\documentclass[letterpaper,11pt]{article}
\usepackage{graphicx,amssymb,amstext,amsmath,amsfonts}
\usepackage{fullpage}
\usepackage{hyperref}
\usepackage{cite}
\usepackage{bm}
\usepackage{comment}                    

\makeatletter
\def\@cite#1#2{{ (#1\if@tempswa , #2\fi)}}
\makeatother

\makeatletter
\def\@biblabel#1{#1.}
\makeatother

\newcommand{\onlinecite}[1]{\hspace{-1 ex} \nocite{#1}\citenum{#1}}

\newcommand{\spacing}[1]{\renewcommand{\baselinestretch}{#1}\large\normalsize}
\spacing{2} 

\renewcommand{\figurename}{{\bf{Fig.}}}

\def\@maketitle{%
  \newpage\spacing{1}\setlength{\parskip}{12pt}%
  {\Large\bfseries\noindent\sloppy \textsf{\@title} \par}%
    {\noindent\@author}%
}

\newenvironment{affiliations}{%
    \setcounter{enumi}{1}%
    \setlength{\parindent}{0in}%
    \slshape\sloppy%
    \begin{list}{\upshape$^{\arabic{enumi}}$}{%
        \usecounter{enumi}%
        \setlength{\leftmargin}{0in}%
        \setlength{\topsep}{0in}%
        \setlength{\labelsep}{0in}%
        \setlength{\labelwidth}{0in}%
        \setlength{\listparindent}{0in}%
        \setlength{\itemsep}{0ex}%
        \setlength{\parsep}{0in}%
        }
    }{\end{list}\par\vspace{12pt}}

\begin{document}

\noindent\parbox{\textwidth}{\center{
{
\LARGE{\textbf{Giant supercurrent states in a superconductor-InAs/GaSb-superconductor junction}}
}}}
\noindent\parbox{\textwidth}{\center{
{
Xiaoyan Shi$^{1*}$, Wenlong Yu$^2$, Zhigang Jiang$^2$, B. Andrei Bernevig$^3$, W. Pan$^{1*}$, S.D. Hawkins$^1$, J.F. Klem$^1$
}}}
\noindent\parbox{\textwidth}{\center
\today
}
\begin{affiliations}
 \item Sandia National Laboratories, Albuquerque, NM 87185, USA
 \item School of Physics, Georgia Institute of Technology, Atlanta, GA 30332, USA
 \item Department of Physics, Princeton University, Princeton, NJ 08544, USA
\end{affiliations}

\noindent\textbf{Abstract}

\noindent{
Superconductivity in topological materials has attracted a great deal of interest in both electron physics and material sciences since the theoretical predictions that Majorana fermions can be realized in topological superconductors \cite{Hasan_Kane_2010,RevModPhys.80.1083,RevModPhys.83.1057,BernevigBook}. Topological superconductivity could be realized in a type II, band-inverted, InAs/GaSb quantum well if it is in proximity to a conventional superconductor. Here we report observations of the proximity effect induced giant supercurrent states in an InAs/GaSb bilayer system that is sandwiched between two superconducting tantalum electrodes to form a superconductor-InAs/GaSb-superconductor junction. Electron transport results show that the supercurrent states can be preserved in a surprisingly large temperature-magnetic field ($T-H$) parameter space. In addition, the evolution of differential resistance in $T$ and $H$ reveals an interesting superconducting gap structure.
}

\vspace{7mm}
\noindent{* Correspondence to: xshi@sandia.gov, wpan@sandia.gov} 

\newpage

Majorana Fermions, which are their own anti-particles, possess non-Abelian properties and could potentially be utilized in topological quantum computations\cite{RevModPhys.80.1083}. It is therefore not surprising that the search for Majorana fermions in solid state systems has attracted a great deal of attention in recent years. Among many theoretical proposals, hybrid systems of an s-wave superconductor in proximity to a 2D topological insulator (TI) with strong spin-orbit coupling\cite{Hasan_Kane_2010,RevModPhys.83.1057,PhysRevLett.95.146802,PhysRevLett.100.096407,127001,PhysRevLett.109.056803} are amongst the most promising. Out of the theoretically known 2D TI systems, HgTe/HgCdTe\cite{bhz,Science76670,PhysRevX.3.021007} and InAs/GaSb\cite{PhysRevLett.100.236601,PhysRevLett.107.136603,1306.1925} are currently under extensive experimental study. In both cases, the conventional semiconductor will have a phase transition to a quantum spin Hall (QSH) insulator when the quantum well (QW) width exceeds a critical value  $d_c$\cite{bhz,PhysRevLett.100.236601,PhysRevLett.78.4613}. In the QSH regime, a 2D TI exhibits an insulating bulk and conductive protected helical edge states. When put in proximity to a conventional s-wave superconductor, it has been shown that superconductivity (which would then necessarily be of topological nature) can be induced in the edge channels. Recent transport experiments in HgTe/HgCdTe\cite{1312.2559} and InAs/GaSb\cite{1408.1701} samples have shown probable edge-mode superconductivity in these 2D topological insulators. However, further studies on supercurrent spatial distributions may be needed\cite{1410.4205}. In this letter, we show the observation of giant supercurrent states in a high quality superconductor-InAs/GaSb-superconductor hybrid device.

\begin{figure}
  \centering
  \includegraphics[width=17cm]{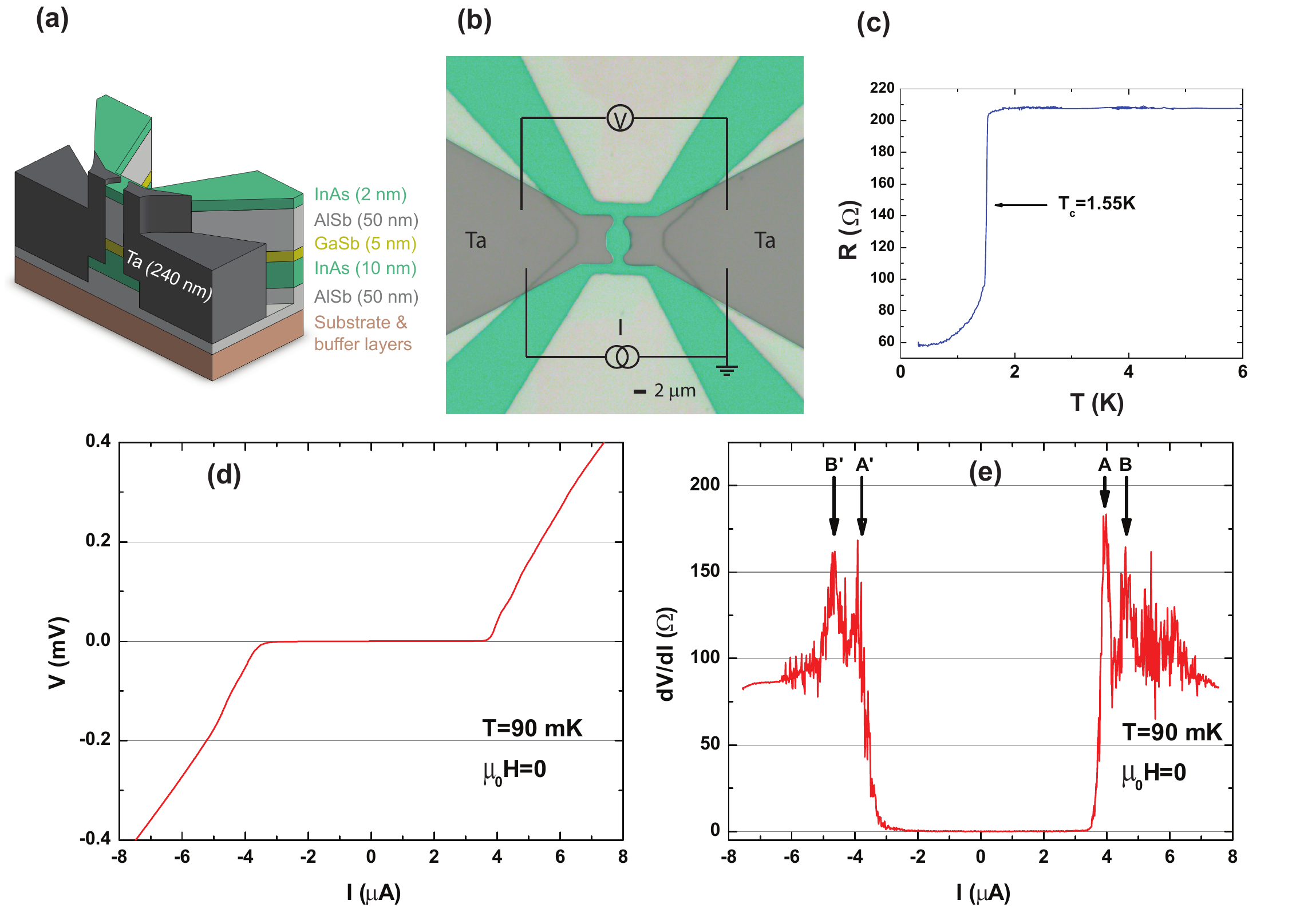}\\
  \caption{\textbf{Sample structure and zero magnetic field transport.} (a) Layer structure of the Ta-InAs/GaSb-Ta junction (not to scale; cutting along the horizontal center line of (b)). (b) Optical image of the sample. The junction is at the center. Two terminal electrical transport measurement setup is overlaid on the image. $I$ represents either a dc current source (for $I$-$V$ measurements) or sum of an ac and a dc currents (for differential resistance measurements). $V$ is either a multimeter or a combination of a multimeter and a lock-in amplifier. (c) $R(T)$ of the junction shows a superconducting transition $T_c\approx 1.5$~K. (d) Typical dc $I$-$V$ and (e) differential resistance \emph{vs}. $I$ traces are shown, respectively, at $T=90$~mK and $\mu_0 H=0$~T. $A$ (or $B$) and $A'$ (or $B'$) mark the position of $\textrm{d}V/\textrm{d}I$ peaks in positive and negative bias, respectively.
  }\label{f1}
\end{figure}

Molecular beam epitaxy techniques were utilized to grow high quality InAs/GaSb bilayer structures. The growth structure of the InAs/GaSb quantum well bilayer is similar to that reported in the past\cite{ApplPhysLett_69_85}, except that the InAs QW width is in the critical regime of $d_c\sim10$~nm (with a GaSb QW width of 5~nm)\cite{InAsGaSb_QW}. An eight-band Kane model calculation shows that at this critical width $d_c$, the lowest electron energy band ($E_1$) in InAs will touch the highest hole energy band ($H_1$) in GaSb at $k=0$ in momentum space, forming a Dirac cone-like feature in the band structure\cite{PhysRevX.3.021007}. Given that the critical width is approximate (even with a realistic eight band Kane model), we can say that the bulk system has a very small (or vanishing) gap near or in the topological regime. Superconducting tantalum (Ta) electrodes were directly deposited onto the InAs layer after photolithography patterning and wet chemical etching procedures to expose that layer. A Ta-bilayer-Ta junction was formed (Fig. \ref{f1}(a) and (b)). The minimum separation of the two superconducting Ta electrodes is $L\approx 2~\mu$m \cite{science_notes}.

\begin{figure}
  \centering
  \includegraphics[width=17cm]{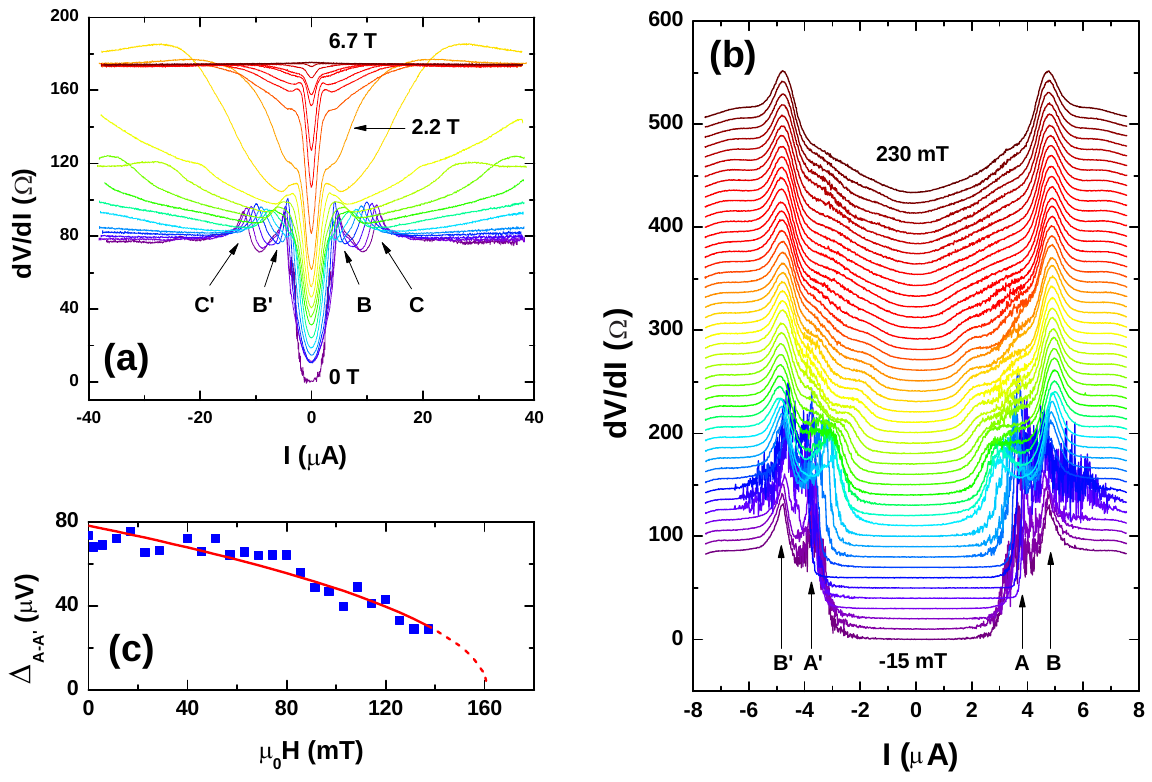}\\
  \caption{\textbf{Magnetic field dependence of {\bm $\textrm{d}V/\textrm{d}I$ on $I$ at $T=90$}~mK.} (a) $\textrm{d}V/\textrm{d}I$ \emph{vs}. $I$ in a large dc $I$ range. From bottom to top, applied magnetic field increases from 0~T to 6.7~T. $B~(B')$ and $C~(C')$ mark peak positions. (b) Detailed plots of $\textrm{d}V/\textrm{d}I$ in small magnetic fields. From bottom to top, the field increases from -15~mT to 230~mT. Curves are shifted vertically for clarity. (c) Voltage difference between peaks $A$ and $A'$ ($\Delta_{A-A'}$) as a function of applied field. Both solid and dashed lines show a phenomenological fit as $\Delta_{A-A'}=c~\left(1-\frac{H}{H_{c1}}\right)^{1/2}$ with $c=78.3~\mu\textrm{V}$ and $\mu_0 H_{c1}=161$~mT.
  }\label{f2}
\end{figure}

Fig. \ref{f1}(c) shows the temperature ($T$) dependence of the junction resistance ($R$) in absence of a magnetic field ($H$). A sharp superconducting transition of the Ta leads can be seen at 1.5~K, with no noticeable change of resistance near 4.47~K (the superconducting transition temperature $T_c$ of Ta bulk). The finite resistance at low temperature is due to a section of the bilayer between the two Ta electrodes. As temperature is further decreased, a dissipationless supercurrent is observed to pass through the junction as shown in Fig. \ref{f1}(d) and (e). A typical dc current-voltage ($I$-$V$) trace at $T=90$~mK is shown in Fig. \ref{f1}(d). For large $I$ ($|I|>20~\mu$A, not shown), the $I$-$V$ curve is simply a straight line indicating a normal state following Ohm's law. As $|I|$ decreases below a critical value $I_c\approx 3.5~\mu$A, the voltage across the junction maintains a zero value and clearly demonstrates a complete superconducting path throughout the junction, which is consistent with the differential resistance ($\textrm{d}V/\textrm{d}I$) result shown in Fig. \ref{f1}(e). In the intermediate $I$ regime between $I_c<|I|\lesssim 8~\mu$A, fluctuations are observed in the $I$-$V$ trace.These fluctuations are clearly seen in the $\textrm{d}V/\textrm{d}I$ curve of Fig. \ref{f1}(e) (see also Fig. \ref{f2}(b) and Supplementary Fig. S1). We notice that these are random temporal fluctuations. Details of the fluctuations in $\textrm{d}V/\textrm{d}I$ are not reproduced in repeated measurements. This is different from the fluctuations observed in strained HgTe\cite{PhysRevX.3.021007}, where aperiodic fluctuations in $\textrm{d}V/\textrm{d}I$ are reproducible in voltage bias for different $\textrm{d}V/\textrm{d}I$ measurements. In Fig. \ref{f1}(e), two pairs of $\textrm{d}V/\textrm{d}I$ peaks (marked as $A$, $A'$ and $B$, $B'$ in graph) can be seen above the fluctuating background. These peaks (or differential conductance dips) near zero bias normally signal the breakdown of Cooper pairs near the critical current $I_c$.

In Fig. \ref{f2}(a), $\textrm{d}V/\textrm{d}I$ is displayed over a large dc bias current range. An additional pair of pronounced $\textrm{d}V/\textrm{d}I$ peaks ($C$ and $C'$) appear at higher bias current. We note that the current positions of all three peaks are reproducible and robust against the random temporal fluctuations but sensitive to external magnetic field. With increasing field, the existing peaks occur at smaller currents. By tracking the field dependence of peaks, we may acquire information about the superconducting gap structures. For example, the $A$ and $A'$ peaks disappear at $\sim 140$~mT (Fig. \ref{f2}(b)). Surprisingly, at the same field, the fluctuations in $\textrm{d}V/\textrm{d}I$ are also greatly suppressed, too. This thus indicates a direct link between the appearance of $A~(A')$ peaks and the $\textrm{d}V/\textrm{d}I$ fluctuations. By plotting the same data in Fig. \ref{f2}(b) as $\textrm{d}V/\textrm{d}I$ \emph{vs}. $V$, where $V$ is the dc bias across the junction (Supplementary Fig. S1), one can define the voltage difference between a pair of peaks, for example, $\Delta_{A-A'}(H=0)=74~\mu$V is for peaks $A$ and $A'$. Fig. \ref{f2}(c) shows the $\Delta_{A-A'}$ evolution in magnetic field based on the data in Fig. \ref{f2}(b). This dependence can be fitted well by a BCS-gap-like phenomenological equation $\Delta_{A-A'}(H)=c~\left(1-\frac{H}{H_{c1}}\right)^{1/2}$ with $c=78.3~\mu\textrm{V}$ and $\mu_0 H_{c1}=161$~mT (Solid and dashed lines in Fig. \ref{f2}(c)). Considering the small energy gap associated with the $A (A')$ peaks and that the $A (A')$ peaks are destroyed by relatively weak magnetic field\cite{Science76670}, here we propose to attribute the $A (A')$ peaks to the onset of the induced superconductivity in the edge channels. This assignment is also consistent with a recent observation\cite{1408.1701}. As shown in a band inverted InAs/GaSb S-N-S junction in Ref.\onlinecite{1408.1701}, the induced edge channel superconductivity is very weak, \emph{i.e.}, it shows up only in a small current ($\sim$nA) and magnetic field ($\sim$mT) region. For $B$ and $B'$ peaks, $\Delta_{B-B'}(H=0)=0.25$~mV and the peaks vanish at $\mu_0H=1.1$~T. For $C$ and $C'$ peaks, $\Delta_{C-C'}(H=0)=1.5$~mV and they vanish at $\mu_0H=2.2$~T (see Supplementary Fig. S2). Assuming that  $\Delta$ corresponds to a superconducting gap, the corresponding critical temperatures are $T_c=0.26$~K, 0.82~K, and 4.9~K for $A$, $B$ and $C$ peaks, respectively. We note that the $C$ peak critical temperature is close to the $T_c$ of superconducting Ta bulk. With a further increase of the magnetic field, the minimum of $\textrm{d}V/\textrm{d}I$ at $I=0$ (zero bias conductance peak) increases and becomes a local maximum at zero bias at $\sim5.7$~T. At even higher fields, only a weak peak at zero bias current remains, which resembles the $\textrm{d}V/\textrm{d}I$ of bilayer itself (Supplementary Fig. S3).

\begin{figure}
  \centering
  \includegraphics[width=17cm]{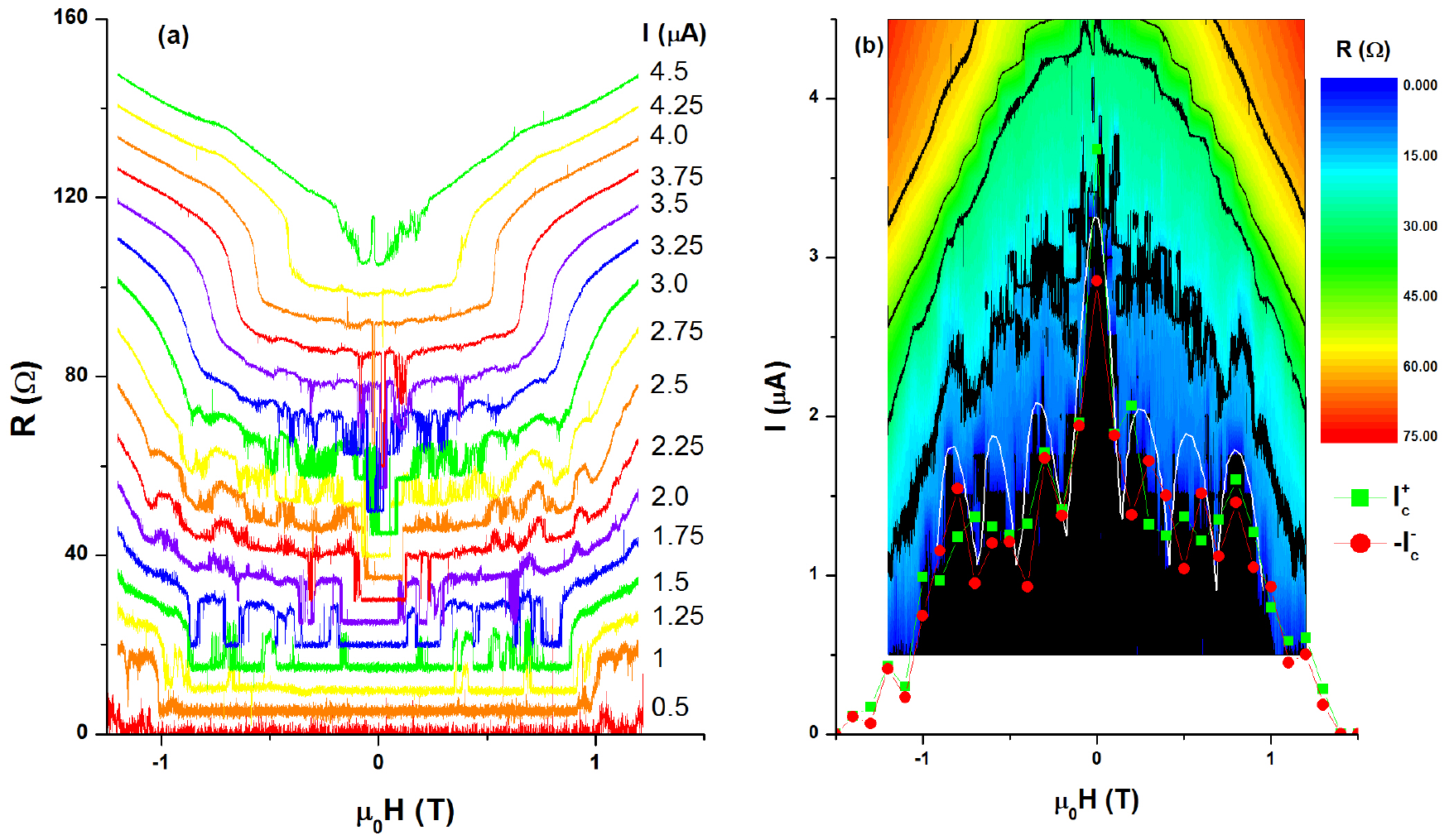}\\
  \caption{\textbf{Magnetic field dependence of {\bm $R$ at $T=30$}~mK.} (a) $R(H)$ traces at fixed dc current. Curves are shifted vertically for clarity. (b) A contour plot of $R(H,I)$ based on data in (a). The lower central black colored area shows the supercurrent region. Critical current values ($I_c^\pm$) from dc $I$-$V$ measurements at given $H$ fields agree with the supercurrent region in the contour plot. $I_c^+$ (green symbols) and $-I_c^-$ (red symbols) are absolute values of the positive and negative critical currents, respectively. A hand-drawn white line outlines the lobes.
  }\label{f3}
\end{figure}

To further investigate the parameter space where supercurrent can exist, dc resistance versus magnetic field traces, $R(H,T=30~\textrm{mK})$ at various DC bias currents ($I$) are shown in Fig. \ref{f3}(a). For the lowest $I$, the superconducting region is surprisingly wide and spans $-1.1<\mu_0 H(T)<1.0$, \emph{i.e.}, the critical field $\mu_0 H_c(T=30~\textrm{mK},I=0.5~\mu \textrm{A})\approx 1$~T. This value is much larger than the critical field of Ta bulk ($\mu_0 H_c(T=0)=0.082$~T, Ref.~\onlinecite{PhysRevB.10.1885}), but consistent with the field where $B$ and $B'$ peaks vanish (Fig. \ref{f2}(a)). Therefore, we attribute $B$ and $B'$ peaks to the induced superconductivity through bulk transport in the bilayer. As the current increases, the critical field $H_c(I)$ decreases as expected. However, some $R>0$ (\emph{i.e.}, non-superconducting) regions show up inside the superconducting regime (Fig. \ref{f3}(a)). This interlace of superconducting and non-superconducting regions indicates a Fraunhofer pattern as normally seen in conventional Josephson junctions\cite{0070648786}. Fig. \ref{f3}(b) displays the same set of data in Fig. \ref{f3}(a) as a contour plot. The black region in the lower central part outlines the superconducting region, which shows a Fraunhofer-like pattern with a central lobe and six side lobes (three on each side). Critical current values  ($I_c^+$ and $-I_c^-$, green and red symbols in the plot, respectively) at fixed $H$ field from another set of dc $I$-$V$ measurements agree with the contour plot. Unlike the standard Fraunhofer pattern, where high order side lobes exist (though their height decays quickly), all higher order ($>3$) side lobes simply vanish in this sample. Moreover, assuming the S-N-S junction area of this sample is about 4~$\mu$m$^2$ (using the area where two electrodes are the closest to each other), $\mu_0H=1$~T field corresponds to about $2000~\Phi_0$ in this junction, or approximately $300~\Phi_0$ per lobe, where $\Phi_0=h/2e$ is the flux quantum. This number is much larger than that of a typical Fraunhofer pattern, which is 1 per lobe for topological trivial superconductivity or 2 per lobe for topological non-trivial superconductivity. In addition, the ratio between the height of the central lobe and the first side lobe is about 1.6, which is in between of 4.7 (for typical Josephson junctions) and 1 (for topological edge state junctions or typical superconducting quantum interference devices, SQUIDs). The zero field resistance value becomes a local maximum for $I>4.4~\mu \textrm{A}>I_c\sim 3.5~\mu\textrm{A}$, which is probably due to a zero field transition to the normal state of the junction. (More curves at this region are shown in Supplementary Fig. S4 for $T=90$~mK.)

\begin{figure}
  \centering
  \includegraphics[width=17cm]{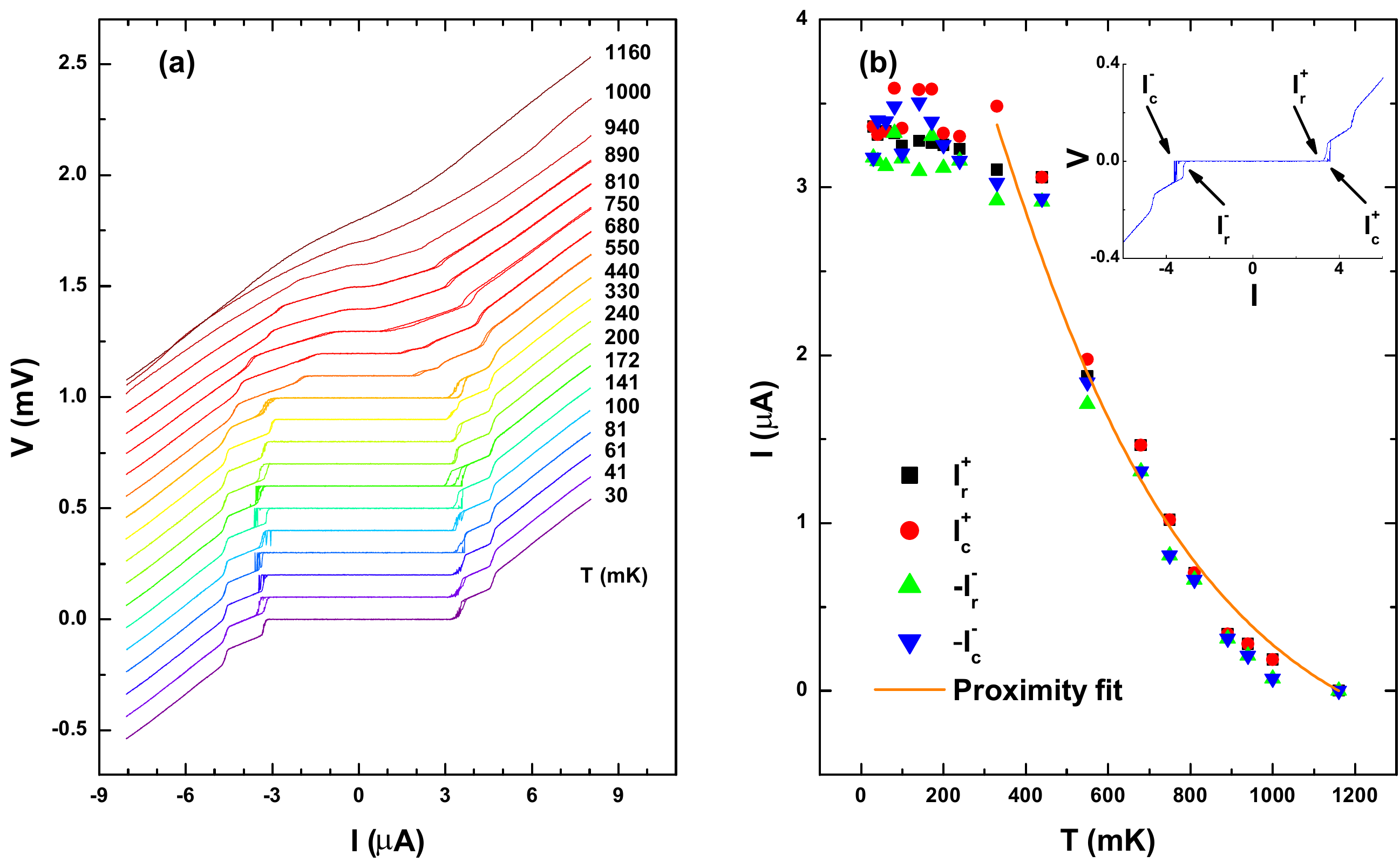}\\
  \caption{\textbf{Temperature dependence of {\bm $I$-$V$} in zero field.} (a) Each $I$-$V$ trace at a given temperature includes both increasing and decreasing current scans. Curves are shifted vertically for clarity. (b) Inset: $I$-$V$ trace at 141~mK is shown as an example to demonstrate the definition of $I_c^\pm$ and $I_r^\pm$, which are positive or negative critical current and retrapping current, respectively. Main plot: Temperature dependence of $I_c^\pm$ and $I_r^\pm$. Solid line shows a fit based on proximity effect for high temperature data ($T\gtrsim 400$~mK). The fitting equation is $I_c \propto \left( \frac{\Delta_N(T)}{\cosh[L/2\xi_N(T)]}\right)^2\frac{1}{\xi_N(T)}$, which gives $T_c=1.16$~K and $\xi_N(T_c)\approx 497$~nm.
  }\label{f4}
\end{figure}

The temperature dependence of $I$-$V$ in zero magnetic field is shown in Fig. 4(a). For each temperature, a dc current is scanned in both increasing and decreasing directions to cover the whole supercurrent region. The superconducting region shrinks as temperature increases and completely vanishes at $T\sim1.16$~K, which we take as the $T_c$ of the induced superconductivity in this junction. Hysteresis loops exist in some traces near the critical current $I_c$. One can trace the temperature dependence of the critical current $I_c$ and the retrapping current $I_r$ (outer and inner boundaries of the hysteresis loop, respectively) as shown in Fig. \ref{f4}(b). In most traces, $I_c$ and $I_r$ are roughly the same, or the hysteresis loop is very small or unobservable. In addition, both $I_c$ and $I_r$ have a weak temperature dependence at $T<400$~mK, then a quick decay at higher temperatures. Data at $T>400$~mK can be fitted by the proximity-effect theory \cite{PhysRevLett.54.2449} as the solid line in the plot. Based on this theory,  $I_c \propto \left( \frac{\Delta_N(T)}{\cosh[L/2\xi_N(T)]}\right)^2\frac{1}{\xi_N(T)}$, where $\Delta_N(T)$ is the induced superconducting gap in the bilayer and $\xi_N(T)$ is the coherence length of the bilayer. Expressions for $\xi_N(T)$ are known in the two limiting cases classified by the ratio of the mean free path in the bilayer, $l_N$, and the superconducting coherence length in the superconductor Ta, $\xi_S$. In the ``clean limit'', \emph{i.e.,} $\xi_S \ll l_N$, $\xi_N(T)=\hbar v_N/2 \pi k_B T$. In the ``dirty limit'', \emph{i.e.,} $\xi_S \gg l_N$, $\xi_N(T)=\sqrt {\hbar v_N l_N /6 \pi k_B T}$. $v_N$ is the Fermi velocity in the bilayer, and $k_B$ is the Boltzmann constant. Measurements in a reference sample (see Methods) show $l_N=822$~nm in the bilayer at $T=0.3$~K. Given $\xi_S=78$~nm in Ta bulk\cite{PhysRevB.10.1885}, our junction is in the clean limit. The fit in Fig. \ref{f4}(b) determines two independent fitting parameters $T_c=1.16$~K and $\xi_N(T_c)\approx 497$~nm, and this agrees well with the value of $\xi_N(T)=480$~nm obtained at $T=T_c=1.16$~K in the clean limit expression. For the discrepancy of this theoretical model and data at the low temperatures, it may be resolved by including the temperature dependence of material related parameters, such as carrier density, into theoretical calculations\cite{PhysRevB.47.2754}.

To complete our electronic transport measurements in our device, we show in Supplementary Figs. S3 and S5 the differential resistance curves of the bilayer itself through a quantum point contact defined by the Ta electrodes. At low temperatures, $\textrm{d}V/\textrm{d}I$ shows hysteresis with changing of the dc current direction. Interestingly, this hysteresis decreases as temperature increases and vanishes at about 400~mK, which is also the temperature where saturation of $I_c(T)$ vanishes (Fig. \ref{f4}(b)). This coincidence could indicate a link between these two different observations.

Measurements of the junction show that the induced superconductivity may exist in both edge channels and the bulk. For edge mode superconductivity, the critical temperature and field are 0.23~K and 140~mT, respectively. The induced superconductivity in the bilayer bulk has a $T_c=1.16$~K, $\mu_0H_c(T=30~\textrm{mK})\approx 1$~T and $I_c(T=30~\textrm{mK})=3.5~\mu$A. Comparing to the superconductivity in bulk Ta, the induced superconductivity is very strong, despite the very large semiconductor region ($L\sim 2~\mu$m) in the junction. Since $L \gg \xi_S $, the Josephson current is expected to be exponentially weak, while it is not the case in this sample. This unusually large proximity effect has been reported in high transition temperature superconductors as the ``giant proximity effect''\cite{Bozovic2004}. To fully understand the aforementioned results in this junction sample, a Hall bar sample with an electrostatic gate was fabricated from the same bilayer wafer as a reference sample. Quantum Hall measurements on this reference sample (Supplementary Fig. S6) show that bilayer wafer sheet resistivity is  0.315~k$\Omega$, the charge carrier density is $n=1.8 \times 10^{11}~\textrm{cm}^{-2}$ and the mobility is $\mu=1.17\times 10^5~\textrm{cm}^2/\textrm{Vs}$ at 0.3~K and zero gate voltage. Given the effective mass of InAs $m^*=0.023~m_e$ ($m_e$ is the free electron mass), one can estimate the Fermi level in the bilayer specimen as $E_F=0.5$~meV or 5.6~K. As for the junction sample, its normal resistance $R_N\sim60~\Omega \ll h/2e^2$, which means the bulk transport is the dominant factor in this QW sample with critical width.

We propose the following possible origins for the greatly enhanced superconductivity induced inside the bilayer. First, we note that the semiconductor-superconductor coupling, or the transparency of the junction, can strongly affect the proximity induced pairing potential in the semiconductor ($\Delta_N$) and $\Delta_N=\Delta_S \times\lambda/(\lambda+\Delta_S)$ (Ref. \onlinecite{PhysRevB.85.064512}). Here, $\Delta_S$ is the pairing potential in Ta and $\lambda$ represents the semiconductor-superconductor coupling. Given the $T_c$ values of the Ta electrodes and the induced superconductivity in the bilayer are 1.5~K and 1.16~K, respectively, the parameter $\lambda$ can be determined from the aforementioned equation, giving $\lambda=3.4~\Delta_S$. This means the junction is very transparent and it is in the strong coupling region\cite{PhysRevB.85.064512}. Therefore, a strong proximity effect is expected. Second, a strong spin-orbit (S-O) interaction can also enhance the $T_c$ as it results in a triplet component in $\Delta_N$ which tends to counter pair breaking effect and stabilize superconductivity\cite{PhysRevB.85.064512}. In this bilayer system, the S-O interaction is strong ($E_{SO}=0.38$~eV for InAs and 0.75~eV for GaSb\cite{PhysRevLett.113.147201}), and is much larger than other energy scales, such as $\Delta_S$ and the Zeeman energy. Experimentally, S-O enhanced superconductivity has been observed in many other systems, such as Pb films\cite{nphys2075}. In addition, the long mean free path inside the bilayer plays an important role. Due to the large separation of the two superconducting electrodes, direct phase coherence of Cooper pairs from different electrodes is very unlikely. However, the bilayer with large $\mu$, $l_N$, and $\xi_N$ provides an ideal medium to enable phase coherence between the two Ta electrodes via the Andreev reflections at the S-N boundaries. Finally, according to the Kogan-Nakagawa mechanism\cite{PhysRevB.35.1700}, $T_c$ could be greatly enhanced by magnetic field in clean samples with long mean free path. Indeed, a field induced increase of induced superconducting gap, or equivalently $\Delta_{B-B'}$, can be clearly observed in the small field region (Supplementary Figs. S1(b) and S2).

In summary, we have observed giant supercurrent states through transport measurements in a superconductor-InAs/GaSb bilayer-superconductor junction with critical QW width. The induced supercurrent states seem to exist both in the edge channels and the bulk of the bilayer. Moreover, the bulk supercurrent states are unexpectedly strong in the temperature-magnetic field parameter space, in contrast to the conventional expectation that only an exponentially weak supercurrent is expected in such a thick junction. New features, such as the occurrence of differential resistance peaks and the fluctuations near the superconducting gap, are interesting and may have important implications in the search for Majorana fermions in superconductor-InAs/GaSb bilayer-superconductor junction devices.

\bibliographystyle{unsrt}

\begin{thebibliography}{10}
\bibitem{science_notes}
Information on materials and methods is available in {S}upplementary {I}nformation.
\end{thebibliography}

\bibliographystyle{unsrt}


\bigskip
\bigskip
\noindent\textbf{Acknowledgements:}
This work is primarily supported by a Laboratory Directed Research and Development project at Sandia National Laboratories. X.S., W.Y., Z.J., and W.P. were also supported by the U.S. Department of Energy, Office of Science, Basic Energy Sciences, Materials Sciences and Engineering Division. Device fabrication was performed at the Center for Integrated Nanotechnologies, a US Department of Energy, Office of Basic Energy Sciences user facility at Los Alamos National Laboratory (Contract DE-AC52-06NA25396) and Sandia National Laboratories (Contract DE-AC04-94AL85000). Sandia National Laboratories is a multi-program laboratory managed and operated by Sandia Corporation, a wholly owned subsidiary of Lockheed Martin Corporation, for the U.S. Department of Energy's National Nuclear Security Administration under contract DE-AC04-94AL85000.

\clearpage

\clearpage

\clearpage

\clearpage


\setcounter{figure}{0}

\makeatletter
\makeatletter \renewcommand{\fnum@figure}{{\bf{\figurename~S\thefigure}}}
\makeatother

\clearpage

\begin{center}
{
\Large{Supplementary information:}\\
\vspace{0.2cm}
\textbf{Giant supercurrent states in a superconductor-InAs/GaSb-superconductor junction}
}
\end{center}

\vspace{0.25cm}
\begin{center}
Xiaoyan Shi, Wenlong Yu, Zhigang Jiang, B. Andrei Bernevig, W. Pan, S.D. Hawkins, and J.F. Klem
\end{center}

\clearpage
\noindent{\textbf{Materials and Methods}}

The InAs/GaSb bilayer sample was grown by molecular beam epitaxy on a GaSb substrate. The thickness of GaSb and InAs quantum well layers was 5.0~nm and 10.0~nm, respectively. The bilayer was sandwiched by two AlSb barrier layers with several buffer layers deposited below and an InAs cap layer above. A mesa of the bilayer was defined by photolithography and wet chemical etching processes. Ammonium hydroxide solution and citric acid/hydrogen peroxide solution were used to selectively etch AlSb/GaSb and InAs layers, respectively. Au/Ti (200/10~nm thick) electrodes were deposited by an e-beam evaporator to connect the InAs/GaSb bilayer at the four corners of the mesa. A second photolithography patterning and wet etching were performed to expose the InAs layer in the center of the mesa, then superconducting Ta (240 nm thick) electrodes are directly sputtered on top of it to form a Ta-bilayer-Ta junction. Au wires were glued by silver epoxy to Au/Ti electrodes and Ta electrodes for measurements. As a reference, a Hall bar sample was made from the same wafer. A gate was fabricated on the reference sample after covering the whole sample by a layer ($\sim$100~nm) of atomic-layer-deposition-grown Al$_2$O$_3$. The carrier density, mobility and their dependence on the gate voltage were measured in this reference sample.

The dc I-V characteristics of the sample were measured with dc voltage or current sources and digital multimeters in a two terminal configuration as shown in Fig. 1(b). For differential resistance measurements, a small ac current ($\sim$~10~nA, 13~Hz) was summed with the dc current then feed to the junction, and the ac voltage response of the sample was measured by the standard lock-in technique.

The sample was cooled to either 90 or 30~mK in two dilution refrigerators. For all measurements in magnetic field, the field directions were perpendicular to the sample surface.

\clearpage

\begin{figure}
  \centering
  \includegraphics[width=15cm]{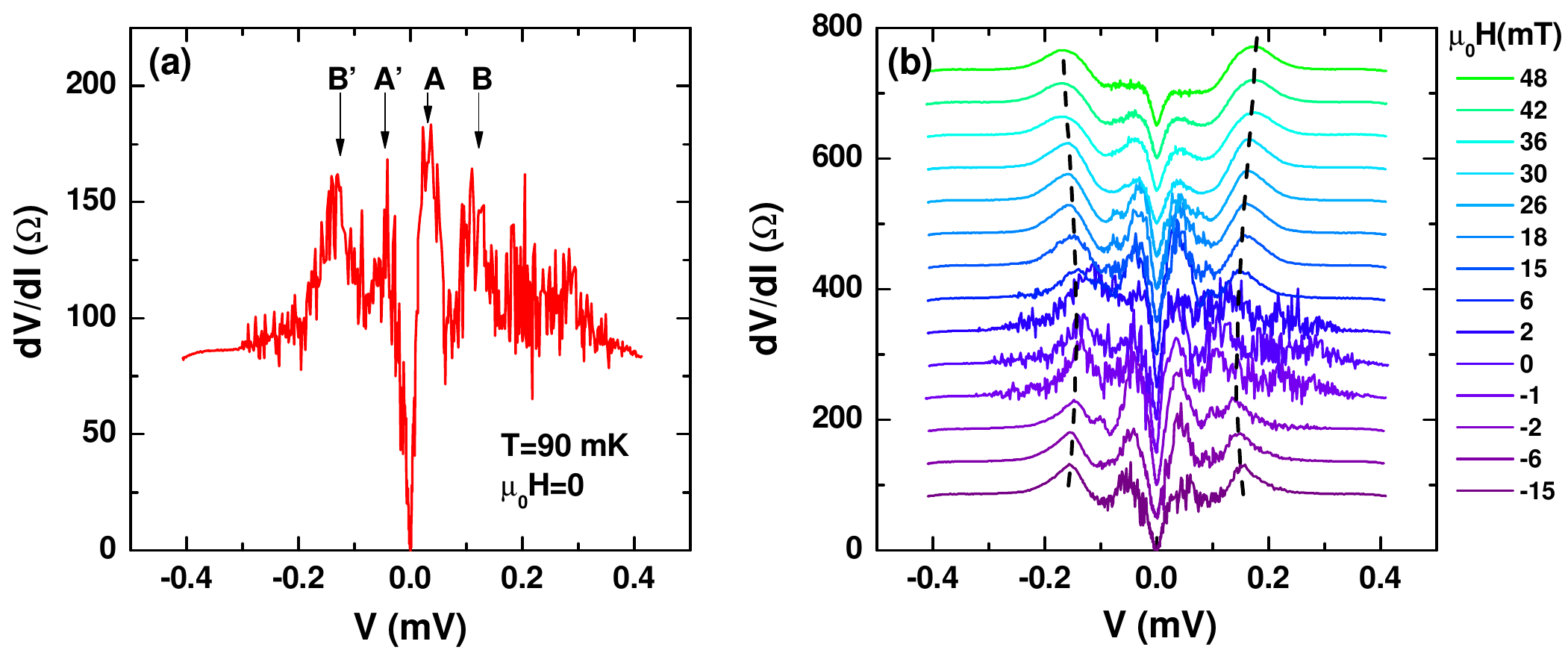}\\
  \caption{\textbf{Differential resistance at {\bm $T=90$}~mK.} (a) $\textrm{d}V/\textrm{d}I(V)$ in zero field. This is the same set of data as shown in Fig. 1(e). (b) Field dependence of $\textrm{d}V/\textrm{d}I(V)$. The same set of data is also shown in Fig. 2(c) as $\textrm{d}V/\textrm{d}I(I)$. A smaller field range is chosen to show details of the fluctuations in $\textrm{d}V/\textrm{d}I$ near zero field. Two hand-drawn dashed lines highlight the $B$ and $B'$ peak positions. As shown, the voltage difference between two peaks, $\Delta_{B-B'}$, increases as field increases in small fields.
  }\label{fs0}
\end{figure}

\clearpage
\begin{figure}
  \centering
  \includegraphics[width=15cm]{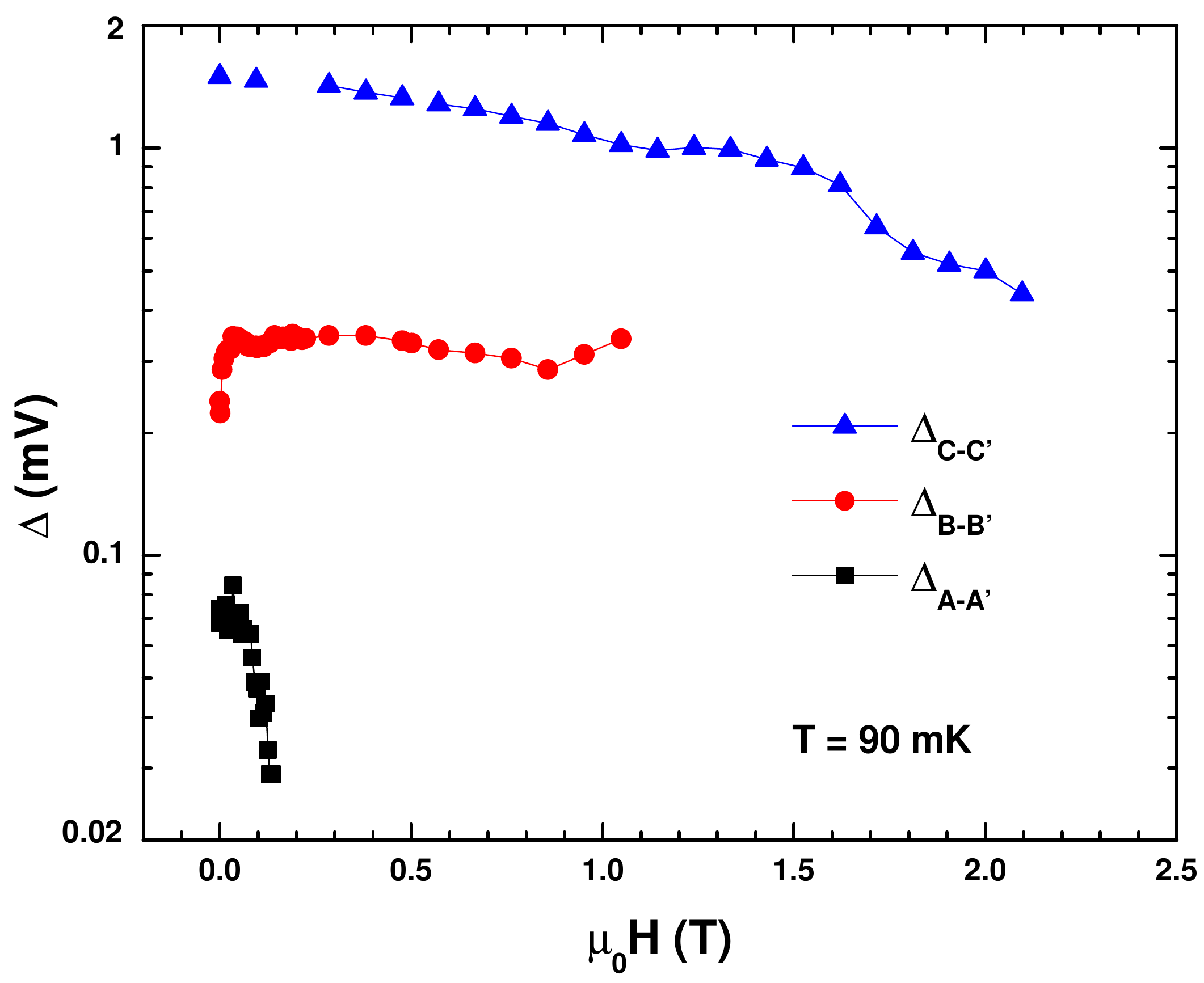}\\
  \caption{\textbf{Evolution of peak separation in magnetic field ({\bm $T=90$}~mK).} $\Delta_{A-A'}$ is the voltage separation for pairs $A$ and $A'$. It decreases monotonically as field increases (also shown in Fig. 2(d)). However, $\Delta_{B-B'}$ increases as field increases initially, then decreases. $\Delta_{A-A'}$, $\Delta_{B-B'}$, and $\Delta_{C-C'}$ are totally suppressed by fields of 0.14~T, 1.1~T, and 2.2~T, respectively.
  }\label{fs02}
\end{figure}

\clearpage
\begin{figure}
  \centering
  \includegraphics[width=17cm]{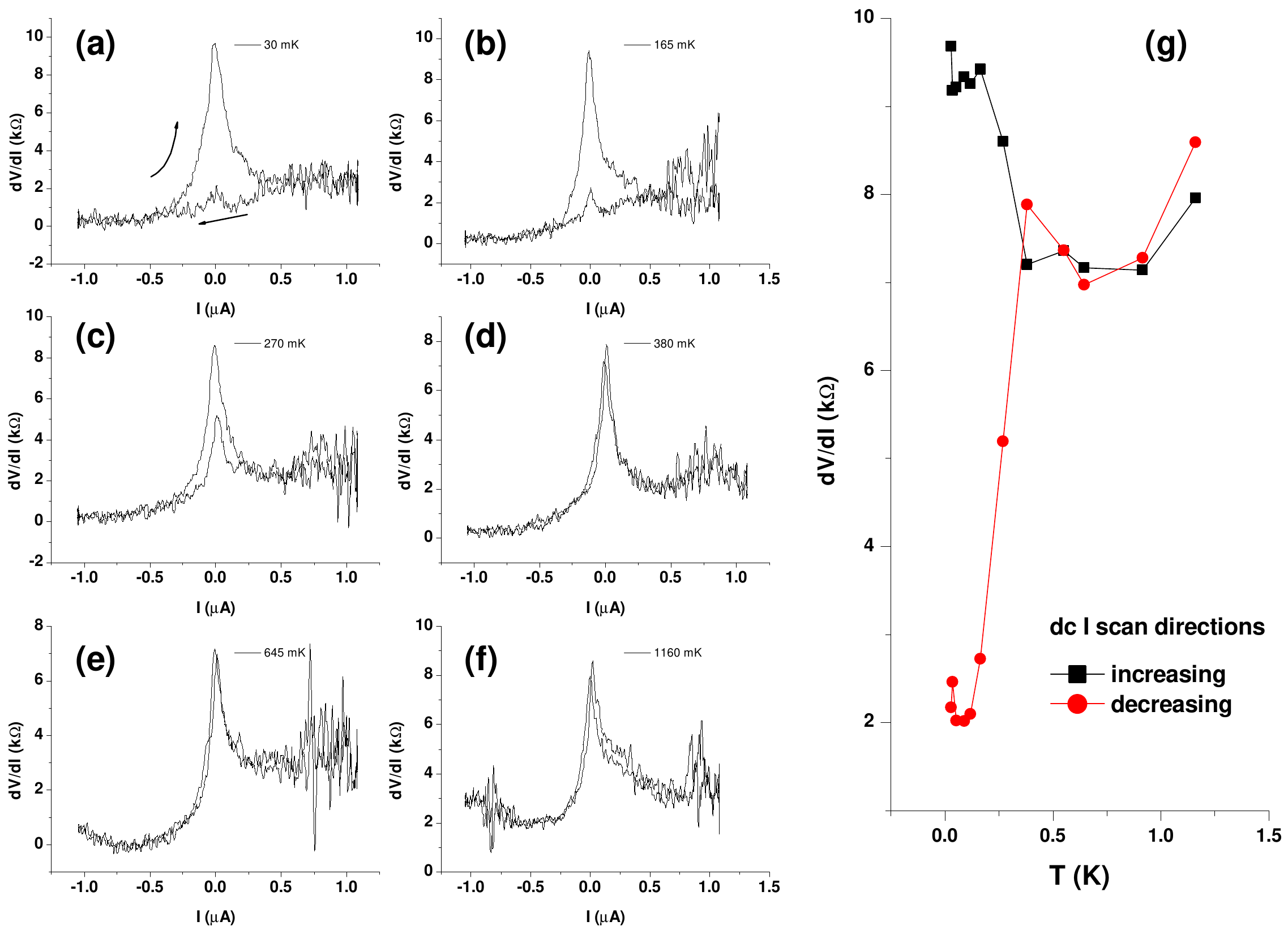}\\
  \caption{\textbf{Temperature dependence of {\bm $\textrm{d}V/\textrm{d}I(I)$} of the bilayer through a quantum point contact defined by Ta electrodes in zero magnetic field.} (a-f) At low temperatures, strong hysteresis exists near dc $I=0$. Arrows represent current scan directions. As $T$ increases, hysteresis decreases and vanishes at $T\approx 400$~mK. The peak values at $I=0$ are shown for different temperatures in (g).
  }\label{fs2}
\end{figure}

\clearpage
\begin{figure}
  \centering
  \includegraphics[width=10cm]{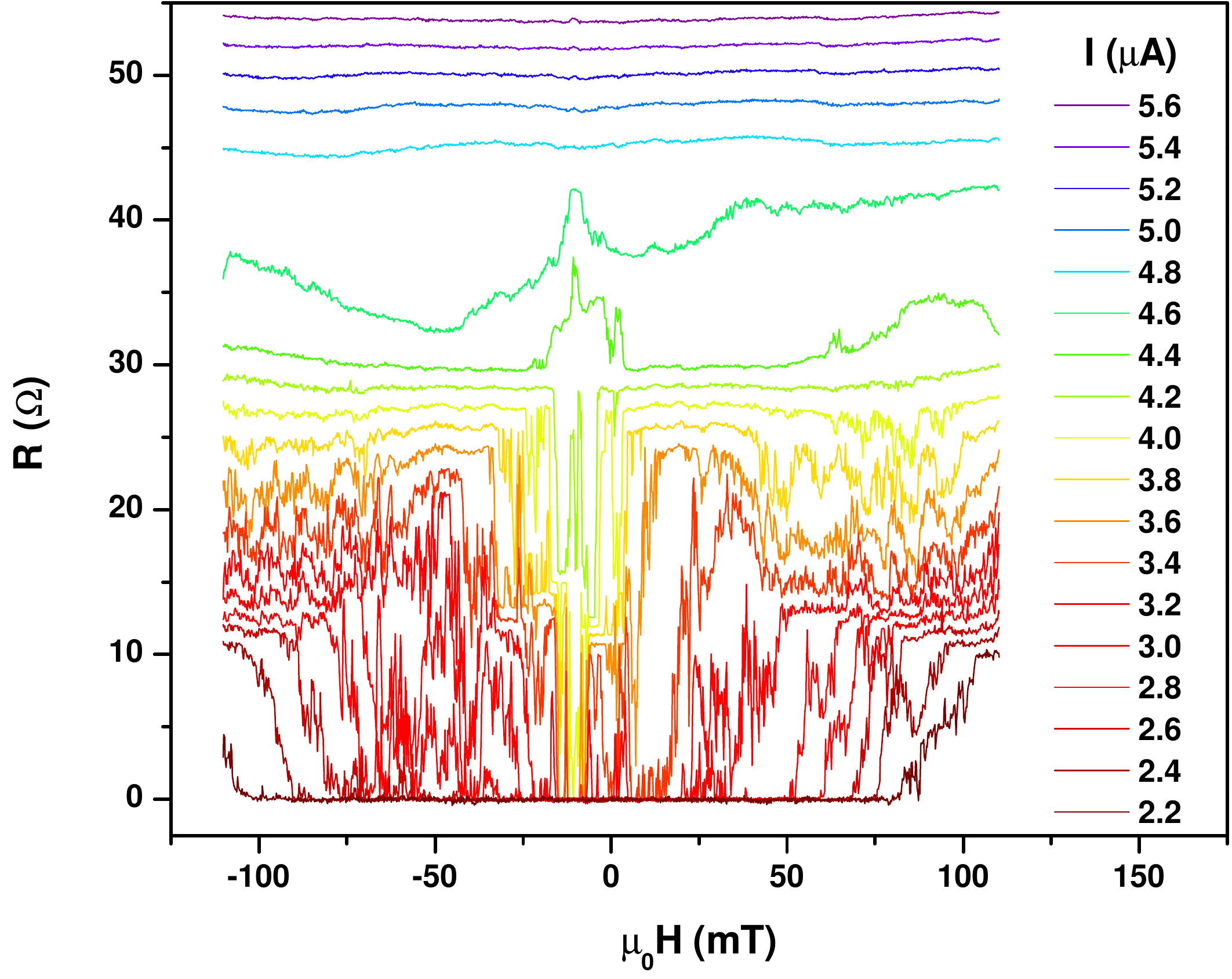}\\
  \caption{\textbf{Magnetic field dependence of {\bm $R$ at $T=90$}~mK.} As current $I$ increases, the zero field superconducting state ($R=0$) observed for $I<4.2~\mu$A is suppressed then turns into a zero field $R$ peak. As $I$ increases further, $R(H)$ becomes a featureless line with high resistance, indicating a normal state with weak magnetoresistance. In addition, pronounced fluctuations of $R$ for low $I$ are shown here.
  }\label{fs05}
\end{figure}

\clearpage
\begin{figure}
  \centering
  \includegraphics[width=15cm]{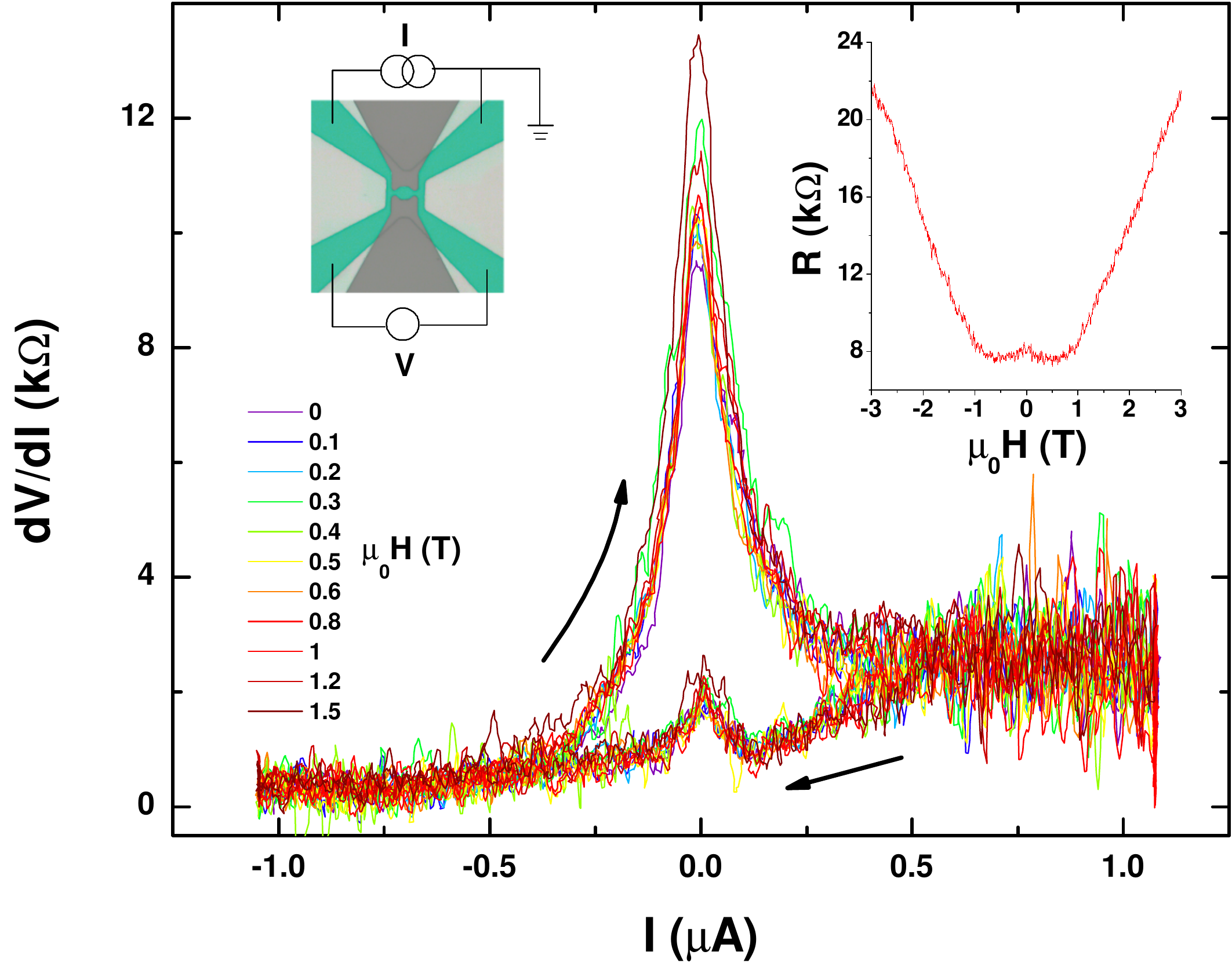}\\
  \caption{\textbf{{\bm $\textrm{d}V/\textrm{d}I(I)$} of the InAs/GaSb bilayer through a quantum point contact defined by Ta electrodes at $T=30$~mK.} The measurement configuration is shown in the left inset. In all magnetic fields, $\textrm{d}V/\textrm{d}I(I)$ shows hysteresis near zero $I$, when the dc bias $I$ scan direction (arrows in main plot) changes. Furthermore, the traces show weak field dependence, except the larger peak near dc $I=0$. The right inset shows the field dependence of the ac resistance $R$ (or differential resistance at zero bias, $\textrm{d}V/\textrm{d}I|_{I=0}$) for the larger peak.
  }\label{fs1}
\end{figure}

\clearpage
\begin{figure}
  \centering
  \includegraphics[width=15cm]{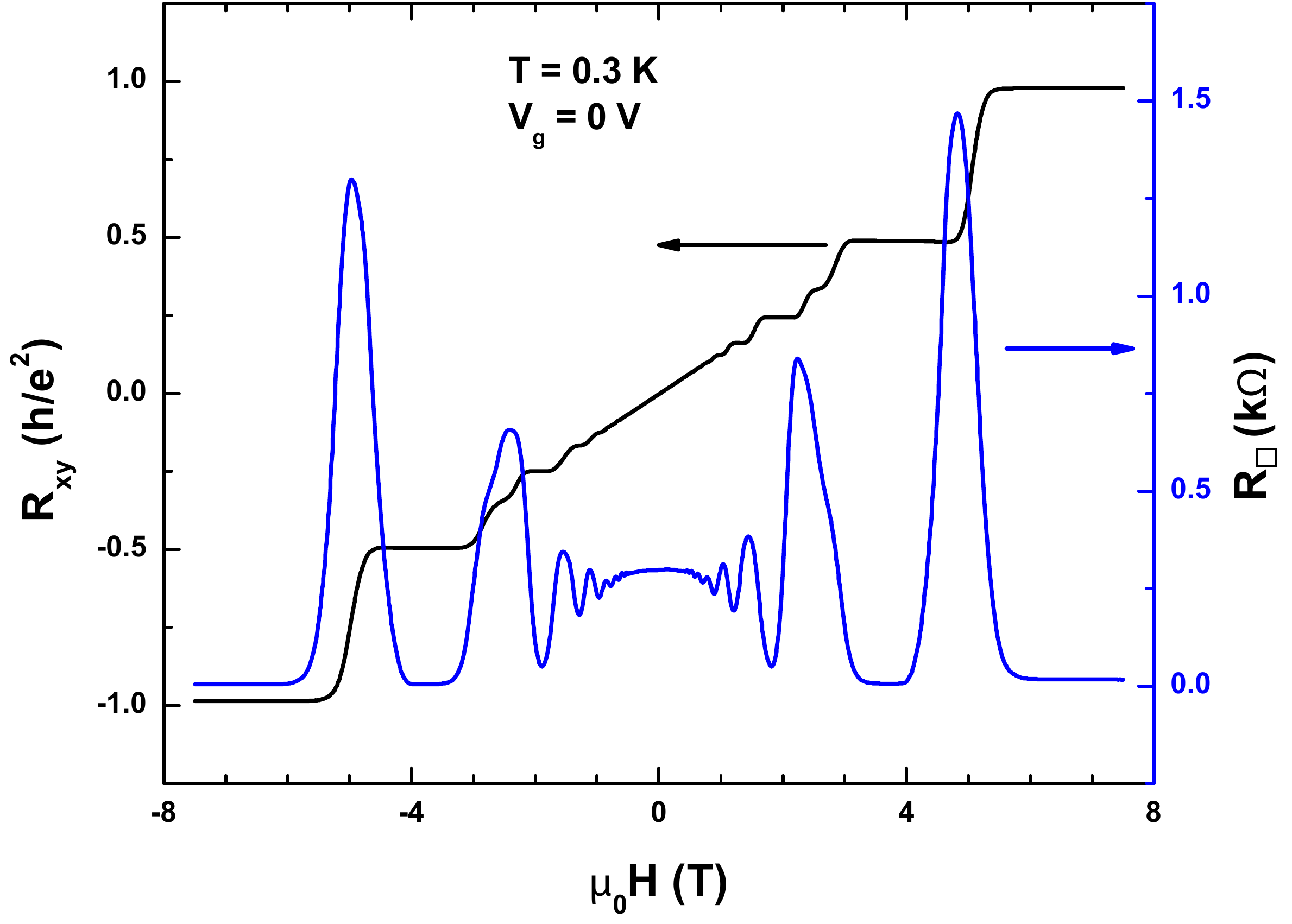}\\
  \caption{\textbf{Quantum Hall measurement of the reference Hall bar sample at {\bm $T=0.3$}~K.} $R_\square$ is the longitudinal sheet resistivity. $R_{xy}$ is the Hall resistance.
  }\label{fs03}
\end{figure}

\end{document}